\documentclass[11pt]{article}
\usepackage{moriond,epsfig}
\usepackage{wrapfig}
\usepackage[justification=justified,singlelinecheck=false]{caption}
\usepackage{array}
\usepackage{amsfonts}
\usepackage{amsmath}

\newcommand{\bsmumu}{$B_{s}\rightarrow\mu^{+}\mu^{-}$}

\newcommand{\bkmumu}{$B^{0}\rightarrow K^{*}\mu^{+}\mu^{-}$}
\newcommand{\bjpsiphi}{$B_s\rightarrow J/\Psi(\mu^+\mu^-)\phi(K^+K^-)$}
\newcommand{\jpsimumu}{$J/\Psi\rightarrow\mu^+\mu^-$}
\newcommand{\lambdappi}{$\Lambda\rightarrow p\pi$}

\def\be{\begin{equation}}
\def\ee{\end{equation}}
\def\bea{\begin{eqnarray}}
\def\eea{\end{eqnarray}}

\begin{document}
\vspace*{4cm}
\title{MUON IDENTIFICATION IN THE LHCb EXPERIMENT}

\author{ X. CID VIDAL, ON BEHALF OF THE LHCb COLLABORATION}

\address{Instituto Galego de F\'isica de Altas Enerx\'ias (IGFAE), Facultade de F\'isica,\\
R\'ua de Jos\'e Mar\'ia Su\'arez N\'u\~nez, s/n, Universidade de Santiago de Compostela,\\ 
15782, Santiago de Compostela, Spain. }

\maketitle\abstracts{
A short summary of the LHCb muon identification procedure is given in this article.
First, the muon system of LHCb is presented, together with some examples of
physics measurements of the experiment where the muon identification is crucial. Then,
the muon identification algorithm is introduced in three single steps. With this, the
efficiency vs. misidentification rate is shown for MC simulated data. The way this method
will be calibrated with real data is also seen. Finally, some preliminary muon
identification results with proton-proton collisions at $\sqrt{s}=900$ GeV are presented.}

\section{Introduction}
LHCb (Large Hadron Collider Beauty) is one of the four major detectors of the LHC (Large Hadron Collider) experiment, that is currently taking place at CERN (Geneva, Switzerland). LHCb will profit from the LHC proton-proton collisions at an energy that is expected to reach 14 TeV in the center of mass to study different aspects of b-physics. Several key measurements of the LHCb experiment rely on the efficient identification of muons. For this purpose, the LHCb detector has a muon system, composed of five muon chambers, and a dedicated algorithm to identify muons. LHCb can identify muons with an efficiency of $\sim95 \%$ and a misidentification rate of $\sim3 \%$, both measured in a set of Monte Carlo (MC) simulated data.

\section{The muon system of LHCb}
\label{sect:muon_system}
\begin{figure}
\begin{minipage}[b]{0.5\linewidth} 
\centering
\includegraphics[width=0.6\textwidth]{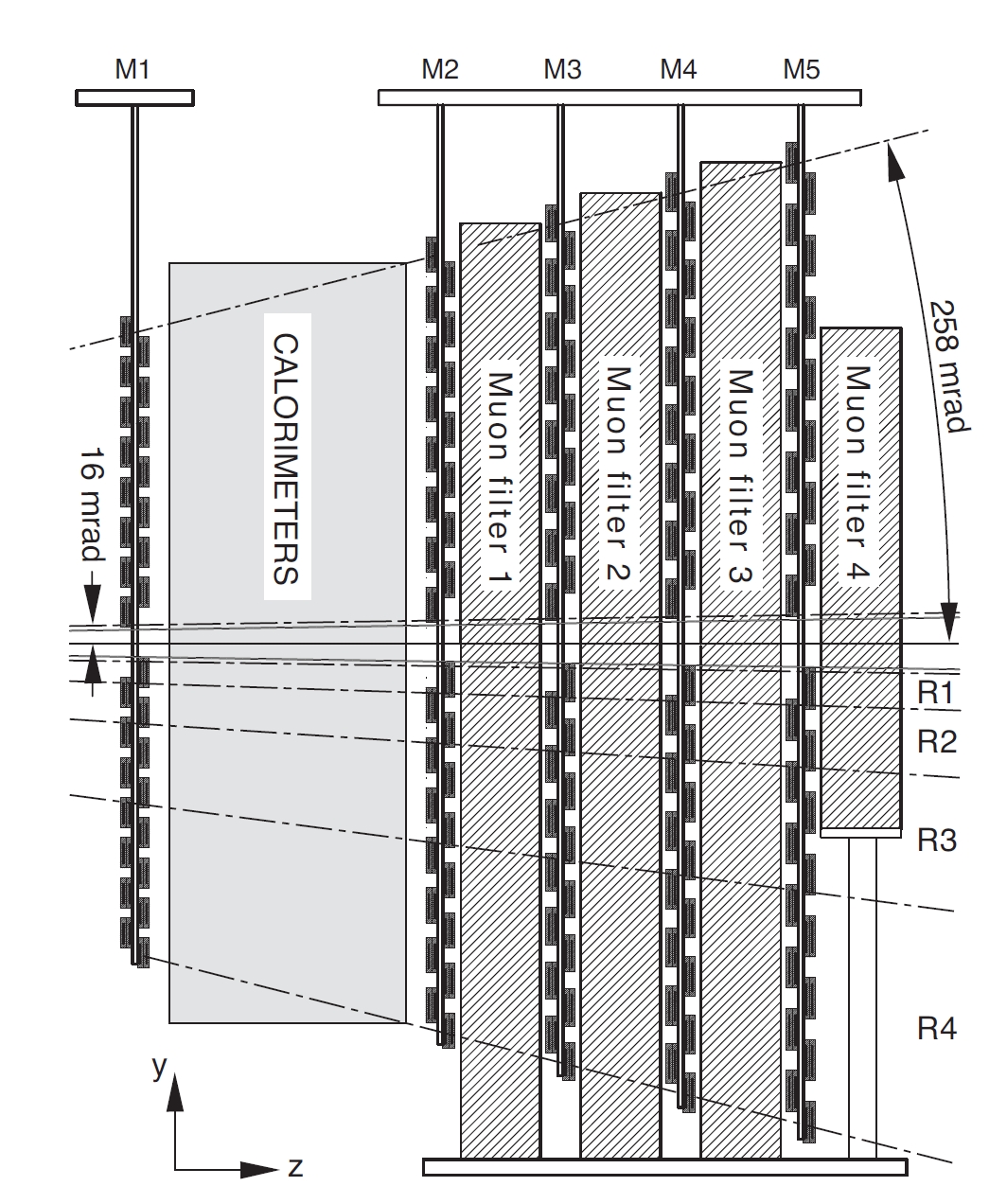}
\caption{Side view of the LHCb muon system,  showing the position of the five stations. The first station is placed before the calorimeters and the other four after them, interleaved with the muon shield.}
\label{fig:muon_system}
\end{minipage}
\hspace{0.45cm} 
\begin{minipage}[b]{0.5\linewidth}
\centering
\includegraphics[width=1\textwidth]{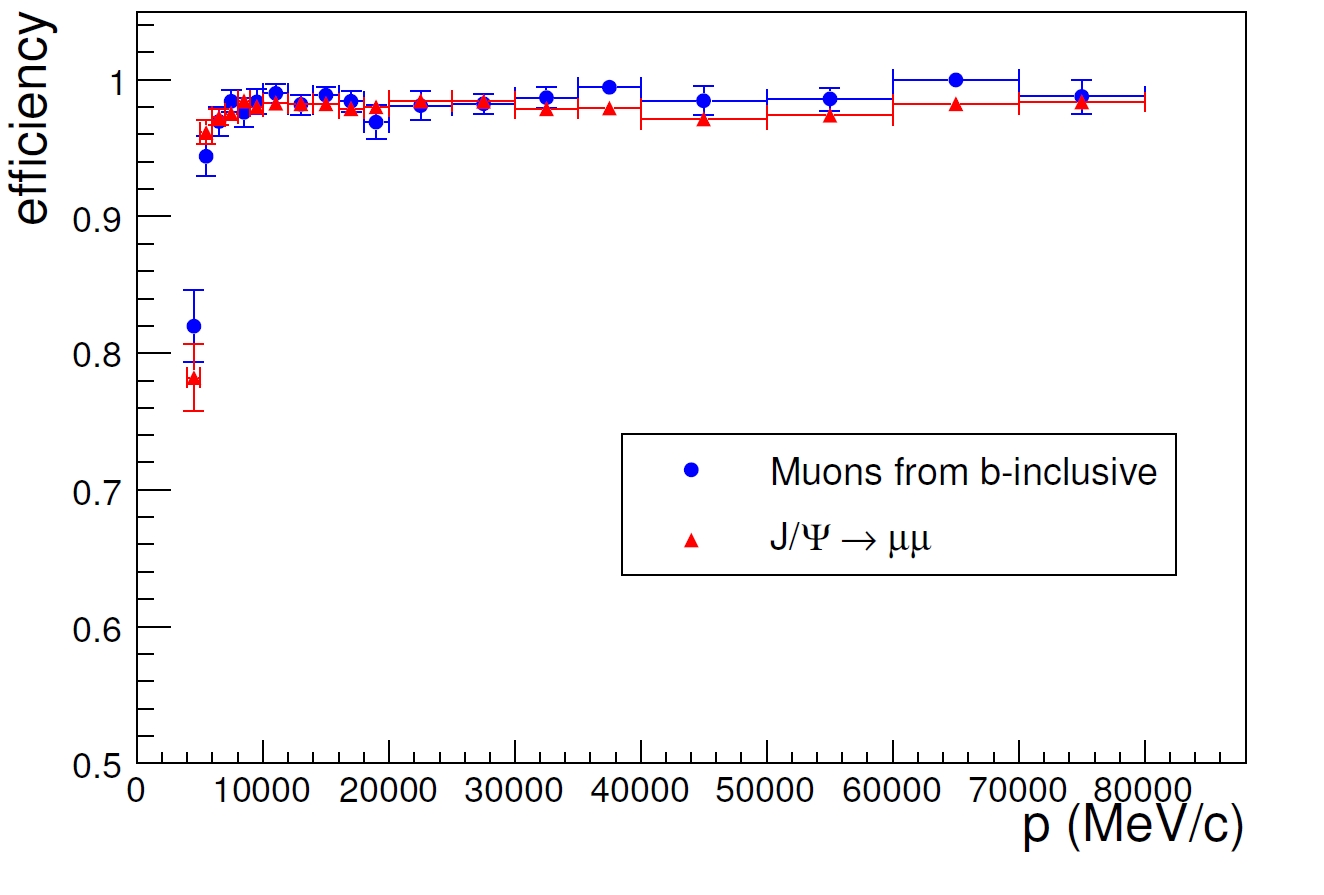}
\caption{Efficiency of the muon candidate requirement vs. momentum for different MC samples. The requirement is calculated on a set of MC Truth matched muons}
\label{fig:muon_candidate}
\vspace{0.5cm}
\end{minipage}
\end{figure}

The muon system of the LHCb experiment~\cite{ref:muon_system} consists of five tracking stations placed along the beam axis and 
separated by iron filters (figure \ref{fig:muon_system}). The first station (M1) is placed in front of the calorimeter preshower, at 12.1\,m from the interaction point, and is important for the transverse-momentum measurement of the muon track used in the first hardware level of the LHCb muon trigger. The remaining four stations are interleaved with the muon shield at mean positions of 15.2\,m (M2), 16.4\,m (M3), 17.6\,m (M4) and 18.8\,m (M5) from the proton-proton interaction point. The shield consists of the electromagnetic and hadronic calorimeters (in front of M2) and three 80\,cm thick iron filters for a total absorber-thickness of 20 nuclear interaction-lengths.

Each station is subdivided in four regions with dimensions and  logical pad size which scales a factor two from one region to the next. 

\section{Physics with muons in LHCb}
LHCb physics interest covers a wide range of topics, from the study of CP asymmetries to the indirect search for New Physics (NP) via the measurement of the branching ratios of very rare decays. Examples of channels of big physics interest and with muons in the final state are \bsmumu, \bkmumu\ or \bjpsiphi.

In particular, the \bsmumu\ channel~\cite{ref:roadmap} is a good example of the potential of LHCb for early discoveries from the proton collisions of LHC. The \bsmumu\ decay takes place through a flavour changing neutral current, and is therefore suppressed in Standard Model at the tree level. In the SM framework, its branching ratio (BR) has been calculated to be $(3.35 \pm 0.32) \times 10^{-9}$, but this value could be very sensitive to the existence of NP, and several models predict an enhancement of this decay. The current best experimental limit is given by the CDF collaboration~\cite{ref:cdf}: $BR(B_s\rightarrow\mu^+\mu^-)< 36 \times 10^{-9}$ at 90\% CL. This is an order of magnitude above SM prediction. The proton-proton collisions at a center of mass energy of 7 TeV in LHC have just begun and this energy is expected to be kept until the end of 2011. With this energy and a small amount of data collected the current limit from CDF is expected to be improved, having by the end of 2010 a sensitivity better than the best from Tevatron experiment (CDF+D0 collaborations) at the end of its lifetime.

\section{The muon identification procedure}
\label{sect:method}
\begin{figure}
\begin{minipage}[b]{0.5\textwidth} 
\centering
\includegraphics[width=0.9\textwidth]{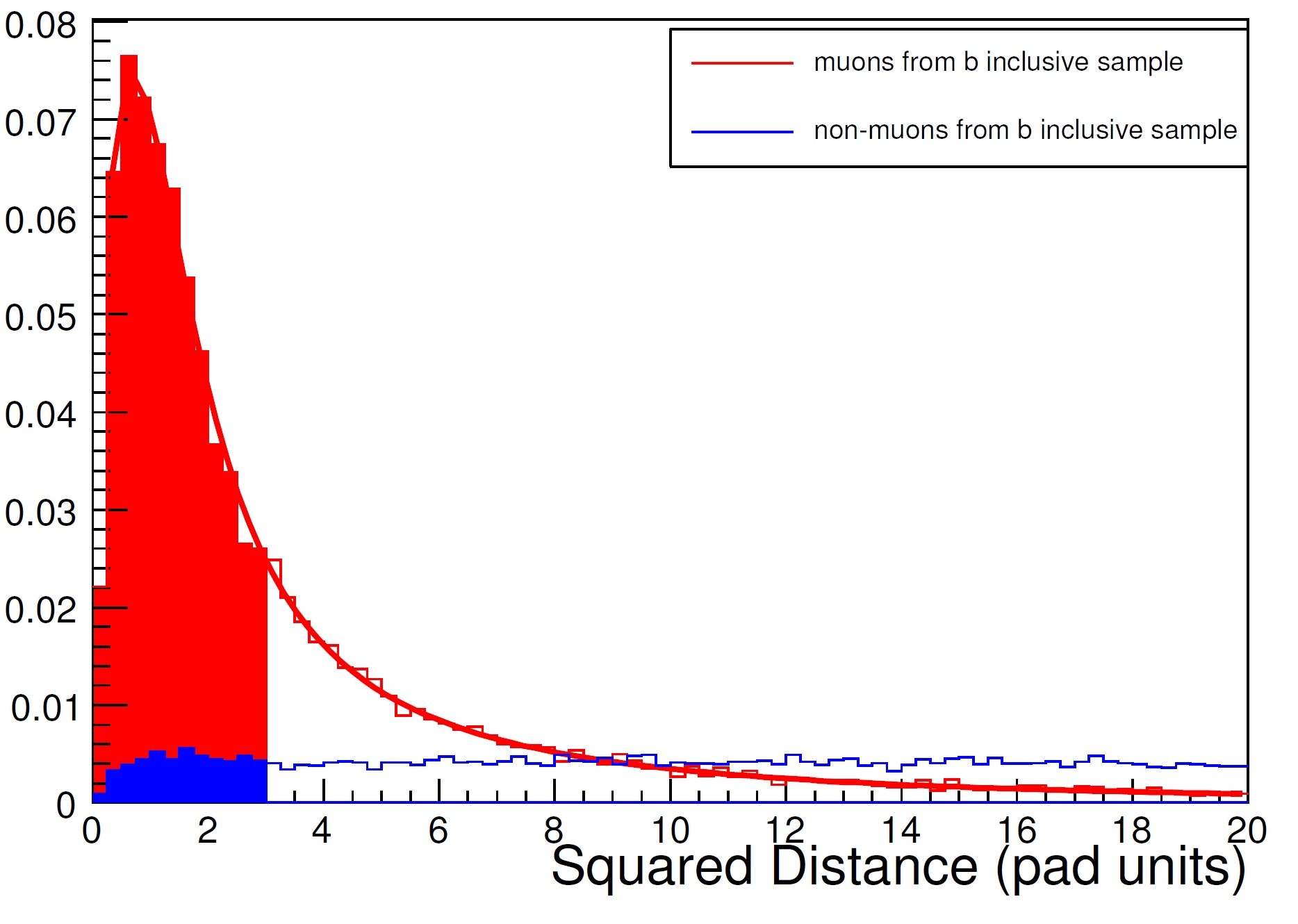}
\caption{\textit{Distance} distribution for muons and non-muon tracks from a MC simulated sample passing the muon candidate requirement. The integrated values shown indicate the muon and non-muon probabilities needed to build the $DLL$.}
\label{fig:dist}
\end{minipage}
\hspace{0.45cm} 
\begin{minipage}[b]{0.5\linewidth}
\centering
\includegraphics[width=1\textwidth]{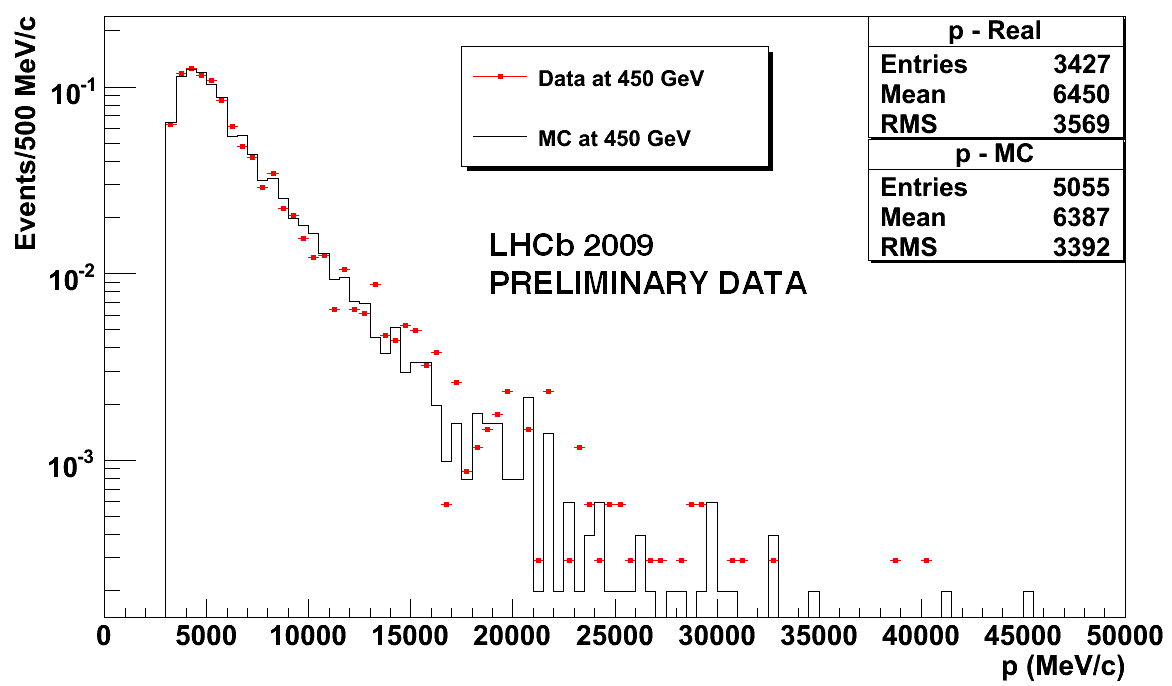}
\caption{Momentum distribution of the tracks passing the muon candidate requirement in 2009 data and comparison to MC simulated data in the same conditions.}
\label{fig:real}
\vspace{1cm}
\end{minipage}
\end{figure}

The tracks built by the tracking system of LHCb can be identified or rejected as muons using the muon subdetector (section \ref{sect:muon_system}). The muon identification algorithm can be divided in three independent steps:
\begin{enumerate}
\item Hits associated to the tracks are looked for in all the five stations of the muon system. To do so, the tracks are linearly extrapolated to the stations and a Field of Interest (FoI) is built around the extrapolated position. This field of interest is different for the $x$ and $y$ coordinates, and depends on the momentum of the tracks, the station and the region of the station where the hits are found.
\item With the identity of the stations with at least one hit in the FoI and the momentum of the tracks, muon candidates are selected. The muon candidate condition requires hits in the FoI of two stations among M2, M3, M4 for tracks with $3<p<6$ GeV/c and in the FoI of three stations among M2, M3, M4, M5 for tracks with $p>6$ GeV/c. This condition removes most of the background (at a $\sim$3-4\% level). The efficiency of this condition vs. momentum in two different MC Truth matched set of muons can be seen in figure \ref{fig:muon_candidate}.
\item Once the muon candidates have been identified, a muon probability can be calculated in order to gain further rejection. A discrimination variable is built with the distance between the position of the extrapolation of the track to each of the muon stations and the closest hit (if any) inside the corresponding FoI averaged to the subdetector's pad size. The exact formula can be found in equation \ref{eq:dist}, where $e_i$ is the extrapolated position, $m_i$ is the position of the hits, $ps$ the size of the pad, $N_{sts}$ is the number of stations with at least one hit in FoI and the sum is done over all stations with at least one hit in FoI. Building the corresponding \textit{distance} distributions for a calibration set of muons and non-muons, allows the calculation of muon ($p_{\mu}$) and non-muon ($p_{non-\mu}$) probabilities (figure \ref{fig:dist}). With the log rate of both these quantities, the Difference Log Likelihood ($DLL=log\left[p_{\mu}/p_{non-\mu}\right]$) will be computed, this being the final discrimination variable of the method.  
\end{enumerate}
\begin{equation}
d^2=\frac{1}{N_{sts}}\displaystyle\sum_{i=x,y}\ \displaystyle\sum_{st=2}^5 \left(\frac{e_i(st)-m_i(st)}{ps_{i}(st)}\right)^{2}
\label{eq:dist}
\end{equation}
With the just described method~\cite{ref:muid_off}, results can be calculated for the efficiencies and misidentification rates provided by different $DLL$ cuts on MC simulated samples. As an example, a $DLL$ cut providing a 94.20\% efficiency on muons would mean a misidentification rate of 3.30\%.

\section{MuonID calibration}
The muon identification calibration with real data requires the use of a set of tracks identified as muons or non-muons without the use of LHCb muon system and with a yield big enough to ensure the monitoring of the method described in section \ref{sect:method}.

As for the muons, the \jpsimumu\ channel can be easily reconstructed in LHCb and will be used to get a rather pure sample of unbiased muons, using a tag and probe method. With two tracks forming a good vertex, passing some $p$ and $p_T$ cuts and with the right $J/\Psi$ mass, the tagged track is required to pass the muon candidate condition, while the probe is required to be a Minimum Ionizing Particle (MIP) in the calorimeter system. The probe tracks are also asked not to have fired the trigger. Once enough probe muons become available, the efficiency of the method will be calculated after a background subtraction. The expected rate for this calibration channel in the 2010 run is above 5 Hz~\cite{ref:muid_calib}.   

The \lambdappi\ channel will be used as a source of non-muons. The very significant expected yield in LHCb allows a very easy reconstruction based simply on kinematical and geometrical cuts. One of the advantages of the \lambdappi\ decay is that the final states, proton and pion, can be easily identified due to the known longitudinal momentum asymmetry between them. This is very interesting for the muon identification measurements, as protons do not decay in flight, while pions do. In the 2010 run, LHCb is expected to reconstruct the \lambdappi\ channel for calibration at a rate bigger than 40 Hz~\cite{ref:muid_calib}. 

\section{MuonID with 2009 data}
The $\sim 6\ \mu b^{-1}$ collected by LHCb in 2009 from proton-proton collisions at $\sqrt{s}=900$ GeV allowed the first results for muon identification with real data. These include mainly the analyses of some distributions for muon candidates and the measurement of the misidentification rate from pions obtained via the $K_s\rightarrow\pi^+\pi^-$ decay. Both results were compared with MC samples simulated with the same data taking conditions. The distributions obtained with muon candidates in data show good agreement with MC simulation. Examples of the variables studied are $p$, $p_T$ and \textit{distance} (equation \ref{eq:dist}). The momentum distributions can be found in figure \ref{fig:real}. As for the misidentification rate, the measured value was $misID(\pi\rightarrow\mu)_{REAL}= (3.8 \pm 0.7) \%$, to be compared with the MC expectation $misID(\pi\rightarrow\mu)_{MC}= (2.3 \pm 0.4) \%$, for tracks in the muon system acceptance with $p>3$ GeV/c. The quoted errors are statistical only.
 
\section{Conclusions}
The muon identification was shown to be crucial for LHCb physics. The muon system was introduced and the identification algorithm, based on three steps, presented. Plans exist on how to calibrate the muon identification algorithm with real data, and a part of these could be already tested with the data collected in 2009 at 450 GeV.

\section*{Acknowledgments}
I would like to thank Laurence Carson, Gaia Lanfranchi and Roger Forty for their kind help in
the preparation of these proceedings.

\end{document}